%% file: paper.tex
\documentclass[lettersize,journal]{IEEEtran}
\usepackage{amsmath,amsfonts}
\usepackage{algorithm}
\usepackage{array}
\usepackage[caption=false,font=normalsize,labelfont=sf,textfont=sf]{subfig}
\usepackage{textcomp}
\usepackage{stfloats}
\usepackage{url}
\usepackage{verbatim}
\usepackage{graphicx}
\usepackage{cite}
\usepackage{tikz}
\usepackage{amsmath}
\usepackage{tikz}
\usetikzlibrary{shapes.geometric, arrows.meta, positioning}
\usepackage{algorithm}
\usepackage{algpseudocode}
\usepackage{amsmath}
\usepackage{booktabs}

\begin{document}

\title{Scalable and Robust \\ Mobile Activity Fingerprinting via \\Over-the-Air Control Channel in 5G Networks}

\author{
    Gunwoo Yoon, and 
    Byeongdo Hong
}

\maketitle

\begin{abstract}
5G has undergone significant changes in its over-the-air control channel architecture compared to legacy networks, aimed at enhancing performance. These changes have unintentionally strengthened the security of control channels, reducing vulnerabilities in radio channels for attackers. 
However, based on our experimental results, less than 10\% of Physical Downlink Control Channel~(PDCCH) messages could be decoded using sniffers.
We demonstrate that even with this limited data, cell scanning and targeted user mobile activity tracking are feasible. This privacy attack exposes the number of active communication channels and reveals the mobile applications and their usage time.

We propose an efficient deep learning-based mobile traffic classification method that eliminates the need for manual feature extraction, enabling scalability across various applications while maintaining high performance even in scenarios with data loss. We evaluated the effectiveness of our approach using both an open-source testbed and a commercial 5G testbed, demonstrating the feasibility of mobile activity fingerprinting and targeted attacks. To the best of our knowledge, this is the first study to track mobile activity over-the-air using PDCCH messages.

\end{abstract}


\section{Introduction}
\label{sec:introduction}
\input{introduction}


\section{Preliminaries}
\label{sec:preliminaries}
\input{preliminaries}

\section{Methodology}
\label{sec:methodology}
\input{Methodology}

\section{Attack}
\label{sec:attack}
\input{attack}

\section{Evaluation}
\label{sec:evaluation}
\input{Evaluation}

\section{Discussion}
\label{sec:discussion}
\input{discussion}

\section{Conclusion}
\label{sec:conclusion}
\input{conclusion}

\bibliographystyle{IEEEtran}
\bibliography{paper}

\newpage

\vfill

\end{document}

%% file: introduction.tex
\IEEEPARstart{I}{n} response to the various privacy threats identified in existing technologies such as Long Term Evolution (LTE), the 3rd Generation Partnership Project (3GPP) undertook extensive discussions\cite{tr33899} and meetings to standardize 5th Generation (5G) technology with enhanced security features\cite{ts33501}. Since the global commercialization of 5G technology, practical privacy vulnerabilities, particularly those related to the Radio Access Network (RAN), have been scarcely reported, indicating that the updated standards effectively mitigate potential attacks. However, unlike LTE, which benefited from the availability of various open-source libraries\cite{srsran, oai, kumar2014lte, bui2016owl, falkenberg2019falcon, hoang2023ltesniffer}, the complexity and bandwidth of 5G initially posed challenges for radio interface analysis. Despite these challenges, with the release of open-source 5G sniffer\cite{5gsniffer}, it has become feasible to conduct analyses through the 5G radio interface.

One of the primary components of 5G networks is the Physical Downlink Control Channel (PDCCH), which plays a critical role in managing data transmission between the base station and User Equipment (UE). The PDCCH carries Downlink Control Information (DCI), which includes scheduling allocations and other control commands essential for network operation. Given its importance, the PDCCH has become a focal point for security studies, particularly in the context of potential threats arising from its analysis.

Various studies in 4th Generation (4G) LTE networks\cite{rupprecht2019breaking, kohls2019lost, bae2022watching, lakshmanan2022privacy, classification_prominent, anamuro2023mobile, baek2023targeted, 9210554} have demonstrated the feasibility of classifying mobile traffic using DCI within PDCCH. These studies employed various machine learning techniques to classify mobile applications based on traffic patterns. However, no similar studies have been conducted in 5G networks yet. The decoding of 5G control channels is more complex compared to LTE, and only recently has a single open-source tool become available\cite{5gsniffer}. Although this tool currently exhibits lower stability and performance compared to LTE tools (Section \ref{subsec:challengesin5Gsniffing}), it marks the first step toward enabling the analysis of 5G mobile traffic classification.

The primary objective of this study is to develop an innovative, efficient, and highly accurate deep learning-based method for mobile traffic classification in 5G networks. Unlike LTE, the 5G PDCCH is dynamically configured using Control Resource Sets (CORESET), enabling more efficient resource allocation and adaptability to varying network conditions. However, this flexibility also presents new challenges in decoding and analyzing control channel information, leading to potential significant data loss during the process of sniffing 5G PDCCH data. To address these challenges, we propose a DCI instance window-based approach that leverages a 1D Convolutional Neural Network (1D-CNN) to analyze the time series data from the PDCCH. This approach eliminates the need for extensive feature engineering and preprocessing, which are typically required by traditional machine learning techniques. Instead, it dynamically adapts to the varying buffering intervals of different applications, providing a robust solution for classifying traffic under conditions of significant 5G PDCCH data loss. This method not only enhances classification accuracy but also strengthens the ability to detect and respond to potential security threats in real-time.

Our study is conducted on standalone (SA) 5G test networks, utilizing real-world applications for video streaming, audio streaming, and Voice over IP (VoIP) calls. This setup enables us to generate and capture 5G mobile traffic, providing a robust dataset for training and evaluating our deep learning model. The results demonstrate that our method achieves high accuracy in classifying mobile traffic across different applications.

The implications of our findings extend beyond traffic classification. The ability to accurately classify mobile traffic in real-time poses significant security risks, as it enables potential attackers to fingerprint user activities and breach privacy. This highlights the need for enhanced security measures in 5G networks, particularly in protecting control channel information. Our research underscores the importance of developing and implementing robust security protocols to safeguard user privacy and ensure the integrity of 5G communications.

In conclusion, this paper introduces a pioneering approach to mobile traffic classification in 5G networks by analyzing the radio interface control channel, leveraging advanced deep learning techniques to enhance both accuracy and efficiency. Our findings contribute to the growing body of knowledge on 5G security and provide a foundation for future research aimed at mitigating the risks associated with this transformative technology. As 5G continues to evolve, ongoing research and innovation will be essential to address emerging threats and ensure the secure and reliable operation of next-generation mobile networks.

The contributions of this paper are fourfold:
\begin{itemize}
\item We propose a novel deep learning model that effectively leverages raw DCI data captured over-the-air from the PDCCH of 5G networks. This model is capable of accurately distinguishing diverse traffic patterns across various applications while maintaining robustness, even in the presence of significant PDCCH data loss during 5G PDCCH sniffing.

\item We conduct extensive experiments on standalone 5G networks with over-the-air traffic using real-world applications. Our evaluation demonstrates the high accuracy and efficiency of the proposed method in classifying different types of mobile traffic.

\item We highlight the security implications of our findings, showing how our traffic classification method can be used to fingerprint user activities and potentially breach user privacy. This underscores the need for improved security measures in 5G networks.

\item To the best of our knowledge, this is the first study to classify applications using DCI data extracted from PDCCH collected over-the-air in 5G networks.
\end{itemize}

In this paper, we particularly focusing on the vulnerabilities associated with the 5G mobile application fingerprinting through 5G radio interfaces. The paper is structured as follows: Section \ref{sec:preliminaries}  provides the necessary background, explores the challenges of 5G sniffing, and introduces related works. Section \ref{sec:methodology} describes the methodology used for data collection and the deep learning model employed for traffic classification. Section \ref{sec:attack} describes the attacks that can be carried out through 5G mobile traffic classification. Section \ref{sec:evaluation} presents our evaluation results in various ways, including the classification accuracy under different parameter changes and methods for acquiring identifiers in 5G. Section \ref{sec:discussion} discusses the feasibility of real-world 5G sniffing, the limitations of our study, and potential countermeasures against the identified privacy risks. Finally, Section \ref{sec:conclusion} concludes the paper by summarizing our findings and suggesting directions for future works.

\subsection{Ethical Consideration}
We conducted all experiments and validations presented in this paper using independently constructed open-source-based and commercial testbeds. During the experimental process, we strictly adhered to privacy protection protocols, ensuring that no personal data was collected. All experiments were conducted using devices and simulation UEs owned by us. Furthermore, this study complied with the General Data Protection Regulation (GDPR), and we provided sufficient information to ensure the reproducibility of the experimental results. Since this study did not involve human subject data, ethical approval was not required.

%% file: preliminaries.tex
This section covers the background knowledge needed for receiving and decoding 5G control messages over the radio interface and explores why radio sniffing is more challenging in 5G than in LTE.
Additionally, this section examines the strengths of our study based on related works and discusses the threat model utilizing our application classification method.

\subsection{Background}

The composition of 5G New Radio~(NR) components\cite{ts38401} helps in understanding the data collection and analysis methods used in this paper. While the control plane architecture of 5G is similar to LTE\cite{ts36401}, it incorporates new technologies and mechanisms, making it somewhat more complex in operation\cite{ts38300}.

PDCCH transmits control information required to schedule data transmission between the base station and UE. The PDCCH includes DCI, which contains scheduling allocations and other control commands that indicate how data should be transmitted and received. Proper functioning of the PDCCH is crucial for efficient network operation, ensuring that resources are allocated correctly and data is transmitted smoothly. While the actual data traffic is encrypted\cite{ts33501}, the PDCCH, which handles resource allocation, remains unencrypted, enabling the monitoring of network traffic. In LTE, the PDCCH has a relatively simple structure\cite{ts36213}, using fixed resource configurations and limited settings. It uses a fixed control resource set to provide control information related to the Physical Downlink Shared Channel (PDSCH). In contrast, 5G NR introduces the concept of CORESET\cite{ts38213}, making the PDCCH structure more flexible and complex. The 5G PDCCH is not fixed and can be dynamically configured using CORESET. This flexibility enhances resource efficiency and adaptability to various network conditions. Additionally, 5G PDCCH employs enhanced scrambling techniques using Radio Network Temporary Identifier (RNTI)\cite{ts38202} to improve security, making the decoding process more complex compared to LTE.

DCI serves as the payload of the PDCCH, containing crucial control information needed for the operation of downlink and uplink channels. DCI includes resource block allocation, modulation and coding schemes, Hybrid Automatic Repeat Request (HARQ) information, and power control commands. This information is essential for the UE to decode data correctly and understand when and how to transmit data. Within the context of the PDCCH, DCI is formatted and encoded as control messages, which are then transmitted via the PDCCH. The base station schedules these messages, and the UE decodes them to extract the control information. The DCI format can vary, providing different types of scheduling information based on network requirements. The flexibility and variety of DCI formats allow efficient control of various traffic types and quality of service requirements.

RNTI is a unique identifier assigned to a UE when it connects to the network. RNTI is used to identify the UE's control and data transmissions within the network. It plays a crucial role in maintaining the security and efficiency of the communication link between the UE and the base station. RNTI ensures that control messages are delivered to the correct UE. All control messages, especially those on the PDCCH, are scrambled with the RNTI before transmission. This ensures that only the intended UE can decode and interpret the messages. By mapping RNTI to specific UEs, the network can manage multiple connections simultaneously, accurately delivering data and control information to the correct recipients. From a traffic analysis and security perspective, RNTI is vital for identifying and segregating individual user traffic, serving as a network management tool and potentially becoming a target for attacks.

\begin{figure}[t]
\centering
\includegraphics[width=0.6\columnwidth]{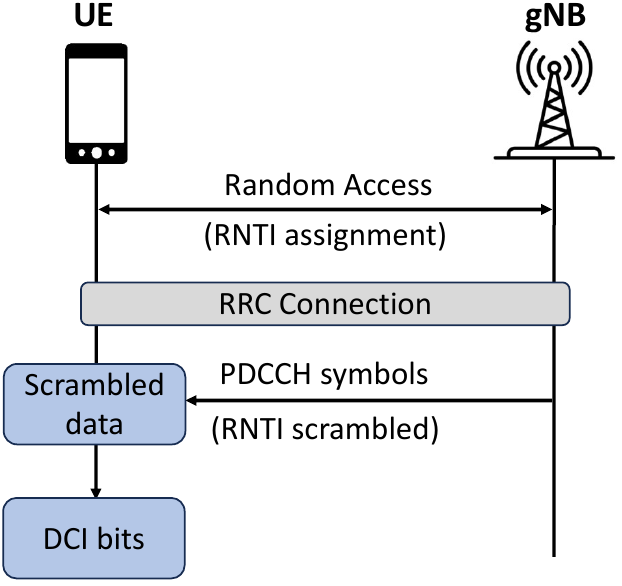}
\caption{Procedure for Obtaining DCI through PDCCH}
\label{fig:dci}
\end{figure}

Figure~\ref{fig:dci} briefly illustrates the process of a next generation Node B (gNB), which acts as the base station in 5G networks, delivering DCI to a UE. First, the UE performs synchronization with the gNB through the \texttt{Random Access} procedure and is assigned a new RNTI before conducting the RRC connection or related settings. Following this, the UE establishes the RRC connection and receives PDCCH symbols from the gNB. 

In both LTE and 5G, the PDCCH symbols are scrambled using the RNTI; however, in 5G, the scrambling-related information, including the RNTI, is transmitted over an encrypted channel.
The UE then uses the RNTI and related information to descramble the data and obtain the DCI bits.

Overall, understanding these components is crucial for comprehending the data collection and analysis methods employed in the paper, as they highlight the advanced mechanisms and flexibility that 5G NR introduces over LTE.

\subsection{Challenges in 5G Sniffing Compared to 4G LTE}
\label{subsec:challengesin5Gsniffing}
Tools such as the UE-side reception tool from srsRAN~\cite{srsran} or Airscope~\cite{airscope} allow for the reception and decoding of 4G LTE PDCCH. However, in comparison to 4G LTE, the structure of 5G is more complex due to additional procedures required for PDCCH decoding on the radio interface. The CORESET utilized in the 5G control channel is mapped to specific time and frequency resources, determining the area where the PDCCH is transmitted in real-time. Unlike 4G LTE, where it was sufficient to observe specific parts of the channel, 5G necessitates real-time adjustments in the observation area. 

Moreover, while both LTE and 5G PDCCH are scrambled using the user's RNTI, in 5G, the RNTI and scrambling-related information are transmitted over an encrypted channel. This difference makes decoding the 5G PDCCH more challenging.

Ludant \textit{et al.} proposed a 5G sniffing method~\cite{ludant20225g, 5gsniffer} that overcomes these challenges by leveraging the low entropy pool of RNTIs and brute force attempts.
Due to the increased complexity of PDCCH acquisition in 5G compared to 4G, we observed a significantly higher rate of missing PDCCH instances when using an open-source 5G sniffer. Since DCI information is embedded within the PDCCH, missing PDCCH instances means missing DCI instances. The number of DCI instances captured by the open-source 5G sniffer was only about 5-10\%~(\S\ref{subsec:datacollection}, \S\ref{subsec:evaluation_in_alternative_commercial_5g_testbeds}) of the total instances generated by the 5G gNB. This high rate of missing DCI data could potentially reduce the accuracy of application classification. However, despite this challenge, we developed and validated an algorithm capable of efficiently and accurately classifying various applications within a 5G network, even under conditions where a significant proportion of DCI data is lost during sniffing.

\subsection{Related Works}
Various methods have been proposed to classify mobile activities based on traffic patterns in LTE networks. Several studies\cite{rupprecht2019breaking, kohls2019lost, bae2022watching, lakshmanan2022privacy} have introduced techniques to distinguish between websites or streaming videos through DCI analysis, but these approaches are based on closed-world assumptions, making it impossible to classify content that has not been previously learned. In the context of mobile activity, the number of applications a person regularly uses is limited\cite{dataprot, buildfire}, whereas the number of websites or streaming videos they access can be significantly more diverse. From the perspective of targeted tracking attacks, it is unrealistic to pre-learn all the content a user might access. Consequently, closed-world-based studies seem to be more effective as investigative tools for identifying prohibited behaviors rather than for targeted security attacks.

An open-world-based analysis method appears necessary for targeted privacy attacks. Classifying mobile applications using an open-world approach enables more realistic targeted attacks. One study focused on manually extracting heuristic statistical metrics to classify applications \cite{classification_prominent}, while others \cite{anamuro2023mobile, baek2023targeted} applied traditional machine learning techniques by manually extracting statistical features. 
Similar to heuristic methods, the performance and scalability of traditional machine learning techniques are expected to decline compared to deep learning, which automatically extracts features as the data increases and the problem becomes more complex ~\cite{goodfellow2016deep, domingos2012few}. Given the vast diversity of mobile applications and the massive volume of data generated in modern mobile networks, manual approaches and traditional machine learning techniques face inherent limitations in distinguishing detailed data characteristics, ultimately restricting scalability. 

Trinh \textit{et al.} demonstrated that deep learning approaches could achieve higher accuracy compared to traditional machine learning methods~\cite{9210554} . However, their classification accuracy significantly dropped when less than 20 seconds of data was secured, and they were unable to extend their approach to targeted security attacks.

One study~\cite{islam2024characterizing} aimed to classify user activity by directly extracting DCI from mobile devices using diagnostic monitor software such as QxDM~\cite{qxdm}. However, collecting control messages via physical connections to the UEs is less practical in real-world scenarios compared to acquiring over-the-air traffic, making these approaches less applicable in realistic environments. Furthermore, the DCI collected from the user-side device differs from the DCI that can be captured over-the-air via control channel messages. In our study, we demonstrated that privacy attacks are feasible by analyzing DCI directly from the radio interface, providing a more realistic and technically sound method.

Ludant \textit{et al}.~\cite{ludant20225g} combined 5G PDCCH sniffing with target traffic generation to determine whether a user was present within the cell. This work is significant in that it demonstrated the possibility of decoding PDCCH messages in 5G. While their privacy attack was limited to confirming the target's presence through simple traffic generation, we extended this by proving that mobile activity tracking is also feasible through PDCCH decoding, beyond just location tracking.

\subsection{Threat Model}
In this paper, we introduce the potential security threats that can arise from analyzing 5G networks, particularly through the radio interface.

\noindent\textbf{Target elements.}~
The primary vulnerable element is the exposure of the privacy of the mobile activities (e.g., Telegram calls, YouTube watching, Spotify listening, Netflix watching, etc.) of users connected to the target 5G network base station.

\noindent\textbf{Attacker model.}~
The attacker is an external entity located within the same base station as the target user and aims to obtain information about the victim's mobile activities.

\noindent\textbf{Attack vectors.}~
The attacker can receive and decode the radio control channel of the 5G base station, specifically the PDCCH. 
Although the user data channel is encrypted, the attacker can analyze data scheduling through the DCI within the PDCCH, allowing them to fingerprint the mobile applications associated with the data stream. 
In this scenario, it is assumed that the victim is using only one application.

\noindent\textbf{Threat scenario.}~
Using existing attack methods~\cite{rupprecht2019breaking, oh2024enabling, ludant20225g}, an attacker can determine the target user's location and RNTI. In this paper, the attack is considered feasible if the target's location can be identified down to the physical cell level. The attacker utilizes the scenario where the target user is connected to the same base station and uses the RNTI to analyze the DCI within the PDCCH. Through this analysis, the attacker can identify which mobile app is associated with a specific data stream. By employing a sniffer, the attacker can continuously monitor mobile app usage activities, leading to a breach of privacy.

%% file: Methodology.tex
In this section, we describe the methodology used for fingerprinting 5G mobile traffic through the radio interface. The methods for collecting, training, and classifying mobile data are detailed.

\subsection{5G Radio Channel Sniffing}
\label{subsec:5g sniffing}
We utilized a free, open-source 5G sniffer~\cite{5gsniffer}. This is the first publicly available free and open-source 5G sniffer. We configured 5G Frequency Division Duplexing~(FDD) mode testbeds to conduct our experiments. Our primary 5G standalone~(SA) FDD mode testbed was established using Open5GS~\cite{open5gs} for the core network and srsRAN for the gNB. Additionally, we set up a secondary 5G SA FDD mode testbed using a commercial Amarisoft solution~\cite{amari_callbox, amari_simbox}. The 5G sniffer was employed to capture PDCCH data over the air, utilizing a Universal Software Radio Peripheral~(USRP) B210~\cite{usrp} to intercept the wireless communication between the UE and the gNB.

\subsection{Training Data Collection}
\label{subsec:datacollection}
We generated 5G mobile traffic using several of the world’s most popular applications for video streaming, audio streaming, and Voice over IP (VoIP) calls. Specifically, for video streaming, we employed four applications: YouTube, Netflix, Disney+, and Prime Video. For audio streaming, we utilized two applications: YouTube Music (YT Music) and Spotify. For VoIP calls, we used WhatsApp and Telegram. The UE in our experiments was a Galaxy S24 Plus, which was used to generate the 5G mobile traffic. The UE was attached to the gNB in our testbed environment, and we assumed that the user would only operate one application at a time, without running multiple applications simultaneously. Additionally, the 5G sniffer captured the over-the-air PDCCH signals transmitted by the gNB to which the UE was connected.

During the collection of 5G mobile traffic from streaming applications, we played random videos and music tracks. In the case of VoIP call traffic, we alternated the UE’s role between receiver and caller to generate DCI data in the PDCCH. We continuously collected DCI data from the PDCCH while the applications were actively running. Over two months, we gathered approximately 713 hours of data using our primary testbed, which is configured with srsRAN and Open5GS, and employed this data to train our deep learning classification algorithm. In this testbed, the ratio of DCI instances captured over-the-air by the 5G sniffer to those generated by the srsRAN gNB was approximately 20:1. This indicates that only about 5\% of the DCI data could be successfully captured over-the-air, with the remainder missed, requiring application classification to be performed using this limited amount of information.

\subsection{Traffic Pattern Characteristics of Mobile Applications.}
\label{subsec:trafficpattern}
To develop a scalable and robust mobile application fingerprinting algorithm, it is essential to understand the specific characteristics of traffic patterns across various applications. For instance, video and audio streaming services like YouTube, Netflix, Disney+, Prime Video, YT Music, and Spotify typically send large bursts of data during buffering periods to ensure smooth playback, rather than streaming data continuously in real-time. These applications transmit burst traffic at intervals influenced by their streaming protocols, network conditions, and the specific application. For example, our measurements show that YouTube typically has burst intervals of 10–15 seconds, Disney+ and Prime Video 5–10 seconds, and Netflix 50–60 seconds. However, these patterns can vary and may fall outside these ranges depending on network conditions. In contrast, applications with VoIP features, such as WhatsApp and Telegram, are characterized by traffic patterns that require low latency and the continuous transmission of small packets, rather than burst traffic. This steady stream of smaller packets is essential for the real-time transmission of voice data. The amount of DCI data generated through the PDCCH is proportional to the actual traffic used by each application. Since this traffic varies over time, it can be classified as time series data. 

\subsection{Training Features}
\label{subsec:trainingfeatures}
The 5G sniffer we used provides DCI instances from PDCCH in raw binary form. We extracted three features from the raw binary DCI data: the direction of Downlink (DL) and Uplink (UL), the Transport Block Size (TBS), and the time intervals between the arrivals of each DCI instance. These features were fed into our deep learning model in their original order of arrival, without any additional preprocessing. This approach eliminates the need for manual feature engineering or statistical feature extraction commonly required in traditional machine learning algorithms, significantly improving efficiency.

\subsection{Classification Strategy}
\label{subsec:classification strategy}

We utilized a 1D-CNN, which is well-suited for analyzing time series data. The 1D-CNN efficiently processes long-term time series data by recognizing patterns along the time axis, effectively capturing both short-term and long-term dependencies through hierarchical feature extraction. 

The input format for the 1D-CNN model is structured as $(n\_timesteps, n\_features)$. Here, $n\_timesteps$ represents the length of the input sequence, specifically the number of data points arranged along the time axis. In our study, each data point corresponds to a single DCI instance.

Next, $n\_features$ refers to the number of features input at each timestep, representing the dimensionality of the data that the model learns from at each timestep. In our case, we extract three features from each DCI instance—direction, TBS, and the time interval between the current and the previous DCI instance, as described in Section~\ref{subsec:trainingfeatures}. 

Based on this input structure, how a single sample is defined for classification or training significantly impacts the model’s performance, efficiency, and accuracy. In a prior deep learning-based study, a sample was defined using a fixed time-based window~\cite{9210554}. This approach gathers DCI instances within a specific time interval and treats that collection as the classification unit. 
However, because PDCCH messages are not transmitted at regular intervals, and the frequency of these captures varies with the application’s data usage, this method requires additional preprocessing, such as interpolation or padding, to standardize input formats with uniform timesteps. 
Moreover, the fixed time window-based method lacks flexibility when managing applications with highly variable traffic patterns. For instance, if the fixed time window is set to 20 seconds, applications like Netflix, which exhibit traffic allocation patterns with intervals around 50 seconds due to buffering, may not have sufficient DCI data captured within that window, potentially leading to inaccurate classification results. 

On the other hand, we define a single sample for classification or training as a \textit{DCI instance window}. This method triggers classification as soon as a predetermined number of DCI instances is collected, without being constrained by the time intervals between them. Since this approach ensures a fixed input length, there is no need for interpolation or padding, allowing for efficient processing. Additionally, since classification is performed only after sufficient DCI instances have been gathered, this method adapts well to applications with varying buffering intervals. This flexibility makes the approach highly scalable, even when new applications with differing traffic patterns are introduced.

Moreover, the DCI instance window method remains robust even in scenarios with high DCI data loss rates. As noted in Section~\ref{subsec:challengesin5Gsniffing}, DCI loss rates can be particularly high when collecting PDCCH data over-the-air in 5G networks. By delaying classification until the required DCI instance window size is reached, the model maintains high accuracy, even in environments with significant DCI data loss.

As described in Section~\ref{subsec:trafficpattern}, streaming applications generate burst traffic in very short periods, typically under 1 second, followed by intervals with little to no data transmission. The occurrence of DCI instances directly corresponds to these bursts. We ensured that the model effectively learns both the burst patterns and the intervals between them by excluding DCI instance windows that are entirely filled by a single burst. This results in a more balanced traffic distribution, enhancing the accuracy and robustness of classification.

\subsection{Training and Classification Model Structure}
\label{subsec:trainingmodel}

Our 1D-CNN model begins with an input layer that processes sequences of DCI data, where the number of timesteps corresponds to the number of DCI instances. The model then progresses through Conv1D layers with ReLU activation and Dropout layers for regularization. It adapts to the length of the sequence using conditional branching: for sequences with 40 or more timesteps, an additional Conv1D layer with 64 filters and a MaxPooling layer is applied; for sequences with 80 or more timesteps, a Conv1D layer with 128 filters, followed by further Dropout and MaxPooling layers, is added. The sequence data is then flattened and passed through a dense layer, culminating in a final softmax output layer for classifying the sequence into one of the possible applications.

\begin{figure}[t]
\centering
\resizebox{\columnwidth}{!}{
\begin{tikzpicture}[node distance=1.2cm, every node/.style={font=\footnotesize}, align=center]
    \tikzstyle{startstop} = [rectangle, rounded corners, minimum width=4cm, minimum height=0.5cm, text centered, draw=black]
    \tikzstyle{process} = [rectangle, minimum width=3.5cm, minimum height=0.5cm, text centered, draw=black]
    \tikzstyle{decision} = [rectangle, rounded corners, aspect=1, minimum width=3.5cm, minimum height=0.5cm, text centered, draw=black]
    \tikzstyle{arrow} = [thick,->,>=stealth]

    \node (input) [startstop] {Input Layer \\ (n\_timesteps, n\_features)};
    \node (conv1) [process, below of=input] {Conv1D Layer \\ 64 filters, kernel size 5 \\ ReLU activation};
    \node (dropout1) [process, below of=conv1] {Dropout Layer \\ 0.2 dropout rate};
    \node (decision1) [decision, below of=dropout1, xshift=-2cm, yshift=0.2cm] {If n\_timesteps $\geq$ 40};
    \node (conv2) [process, below of=decision1] {Conv1D Layer \\ 64 filters, kernel size 7 \\ ReLU activation};
    \node (dropout2) [process, below of=conv2] {Dropout Layer \\ 0.2 dropout rate};
    \node (pool1) [process, below of=dropout2] {MaxPooling1D Layer \\ Pool size 2};
    
    \node (decision2) [decision, below of=dropout1, xshift=2cm, yshift=0.2cm] {If n\_timesteps $\geq$ 80};
    \node (conv3) [process, below of=decision2] {Conv1D Layer \\ 128 filters, kernel size 9 \\ ReLU activation};
    \node (dropout3) [process, below of=conv3] {Dropout Layer \\ 0.3 dropout rate};
    \node (pool2) [process, below of=dropout3] {MaxPooling1D Layer \\ Pool size 2};
    
    \node (flatten) [process, below of=pool2, xshift=-2cm] {Flatten Layer};
    \node (dense) [process, below of=flatten] {Dense Layer \\ 256 units \\ ReLU activation};
    \node (dropout4) [process, below of=dense] {Dropout Layer \\ 0.1 dropout rate};
    \node (output) [startstop, below of=dropout4] {Output Layer \\ n\_outputs units \\ Softmax activation};

    \node (explanation) [below of=output, xshift=1.6cm] {*n\_timesteps = DCI instance window size};

    \draw [arrow] (input) -- (conv1);
    \draw [arrow] (conv1) -- (dropout1);
    \draw [arrow] (dropout1) -- (decision1);
    \draw [arrow] (decision1.south) --  node[anchor=east]{yes} (conv2);
    \draw [arrow] (conv2) -- (dropout2);
    \draw [arrow] (dropout2) -- (pool1);
    \draw [arrow] (pool1.east) -- (decision2.west);
    \draw [arrow] (decision1.west) -| node[anchor=north west] {no} ++(-0.7,-4.3) |-(flatten.west);
    
    \draw [arrow] (decision2.south) -- node[anchor=east] {yes} (conv3);
    \draw [arrow] (conv3) -- (dropout3);
    \draw [arrow] (dropout3) -- (pool2);
    \draw [arrow] (decision2.east) -| node[anchor=north east] {no} ++(0.7,-4.3) |- (flatten.east);
    
    \draw [arrow] (pool2) -- ++(0,-0.6) -| (flatten);
    
    \draw [arrow] (flatten) -- (dense);
    \draw [arrow] (dense) -- (dropout4);
    \draw [arrow] (dropout4) -- (output);
\end{tikzpicture}
}
\caption{Classification Model Used in This Paper}
\label{fig:cnn_model}
\end{figure}
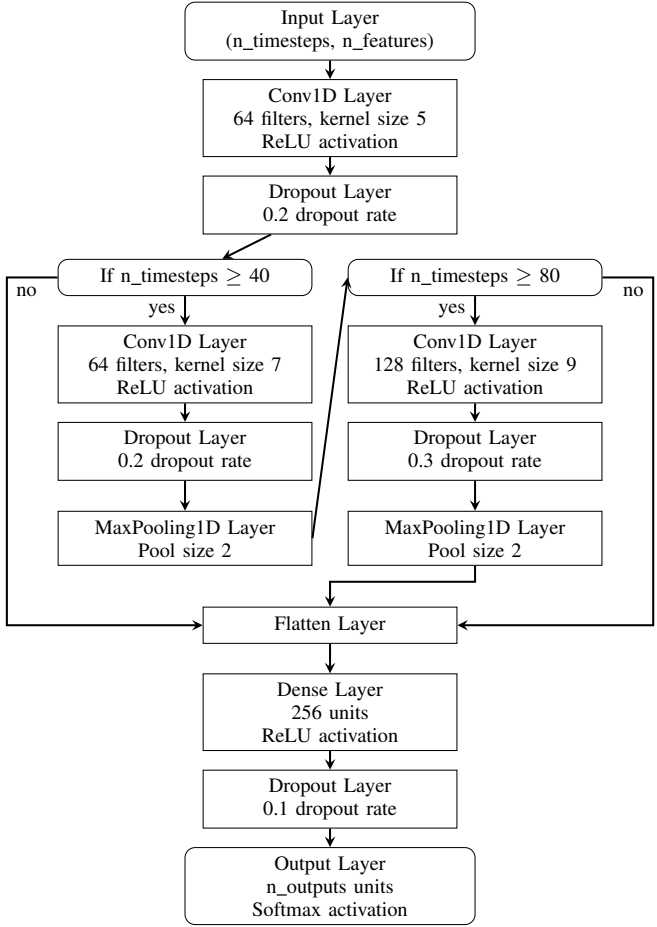

To optimize hyperparameters such as dropout rate, kernel size, and filter size, we employed KerasTuner~\cite{kerastuner}, which dynamically adjusts these parameters within a predefined range. We further refined these values experimentally. The final model architecture is depicted in Figure~\ref{fig:cnn_model}.

%% file: attack.tex
We have categorized attacks through 5G PDCCH decoding into cell scanning and targeted attacks. Cell scanning is a prerequisite for conducting targeted attacks, thereby making these two types of attacks interdependent. In this section, we describe methods and results for an attack aimed at fingerprinting both all mobile devices connected to a cell and the mobile applications of a specific target user.

\subsection{Target Identification within the Cell}
\label{subsec:target identification}

To conduct an attack on a target user, it is necessary to identify the user within a physical cell. Various studies~\cite{kune2012,altaf,hong2018guti,rupprecht2019breaking,hussain2019privacy,ludant20225g} have already introduced methods for obtaining the identifiers of target users. Since most of these identifier acquisitions were performed in LTE networks, it is essential to evaluate their feasibility in 5G networks from various perspectives.

The identifier required for fingerprinting the traffic of a target user is the RNTI. Early studies~\cite{kune2012,altaf,hong2018guti} related to identifiers mostly aimed at acquiring TMSI, an upper-layer identifier. Rupprecht \textit{et al.}~\cite{rupprecht2019breaking} demonstrated that once the TMSI is obtained, it can be mapped to the RNTI within the \texttt{RRC Connection Setup} signaling. As mentioned in Section~\ref{sec:preliminaries}, this RNTI becomes invalid once the device's RRC connection is disconnected and is reassigned during the next RRC connection. If an attacker has acquired the target's TMSI, they can obtain the newly assigned RNTI through continuous monitoring of \texttt{RRC Connection Setup}. If the attacker has not acquired the TMSI, they can force the target's RRC connection to remain active using the modified scheduling signaling attack~\cite{oh2024enabling} proposed by Oh \textit{et al}. If the target's RRC connection does not disconnect and the RNTI reassignment does not occur, the attacker can continuously utilize the target's RNTI.

\subsection{Attack Procedure}
A sniffer located within the cell performs decoding of all PDCCH channels. Utilizing the methods described in Section 3, it is possible to perform mobile application fingerprinting for each RNTI. This procedure constitutes cell scanning. Following cell scanning, a targeted attack can be executed by identifying the target RNTI. By employing the methods outlined in the Section \ref{subsec:target identification}, preliminary information about the target user and RNTI mapping can be achieved. Once the RNTI is identified, the subsequent step involves identifying the mobile application associated with that RNTI, which is the method used for targeted attacks. Baek \textit{et al}.~\cite{baek2023targeted} introduced privacy attacks that can be executed when successfully tracking the target's traffic. These attacks demonstrated the ability to continuously identify the target's location and mobile activity by utilizing multiple sniffers. Additionally, they showed that it is possible to track other users who are engaged in data communication with the target user.

\subsection{Attack Results}
The privacy information exposed when it is possible to fingerprint the mobile activity of arbitrary users and specific users connected to the base station through the attack procedures are as follows.

\subsubsection{Cell Scanning}
5G PDCCH channel decoding reveals the number of real-time radio connections and their connection durations. Mobile activity fingerprinting of each RNTI discloses the specific mobile applications being used by each user connected to the base station. The combination of these results can provide a real-time overview of the status of all users currently connected to the cell. For instance, at a specified moment, it can indicate which applications, such as YouTube or Spotify, each RNTI is using, as well as the start and end times of these applications for each RNTI.

\subsubsection{Targeted Mobile Activity Tracking}
After obtaining the cell and RNTI of a specific user, a targeted privacy attack becomes feasible. This attack remains effective as long as the target user is within the observable base station. While the 5G sniffer is active, the attacker can continuously monitor the times the target user uses their mobile device, the applications they use, and whether they are currently within the range of the base station. Although the RNTI is usually updated when the radio connection is refreshed, the scheduling attack\cite{oh2024enabling} proposed by Oh \textit{et al}. allows for the fixation of the RNTI for the target user. 
3GPP decided not to apply integrity protection to messages below the PDCP layer in LTE\cite{tr33821}, and this decision has also been extended to 5G\cite{ts33501}. This enables RNTI fixation attacks in 5G, ultimately allowing for continuous privacy attacks on the target.

%% file: Evaluation.tex
We evaluate our method's classification performance using various DCI instance window sizes, focusing on accuracy and classification time. Metrics such as precision, recall, F1-score, and accuracy are used, with experiments conducted in both open-source and commercial 5G testbeds to assess robustness. Additionally, we demonstrate a method to identify target users through crafted traffic signatures, further highlighting the potential for targeted attacks in real-world scenarios.

\subsection{Impact of DCI Instance Window Size on Application Classification Accuracy}
\label{subsec:classfication_accuracy}
We evaluated the accuracy of application fingerprinting by varying the DCI instance window size required for classification. The DCI instance window size refers to the number of DCI instances used to fingerprint an application in a single classification decision. To ensure an unbiased evaluation, we utilized an unseen dataset collected during a different period from the training data, avoiding any overlap with the training phase. The analysis began with a window size of 20 DCI instances and increased in increments of 20, concluding at 160 instances.

We used precision, recall, F1-score, and accuracy as evaluation metrics. Precision is the ratio of correctly predicted positive observations to the total predicted positives,\footnote{$\text{Precision} = \frac{TP}{TP + FP}$, where $TP$ is true positives and $FP$ is false positives.} indicating the accuracy of the positive predictions made by the model. Recall, also known as sensitivity, is the ratio of correctly predicted positive observations to all the observations in the actual class,\footnote{$\text{Recall} = \frac{TP}{TP + FN}$, where $FN$ is false negatives.} measuring the ability of the model to find all the relevant cases within a dataset. The F1-score is the harmonic mean of precision and recall,\footnote{$\text{F1-score} = 2 \times \frac{\text{Precision} \times \text{Recall}}{\text{Precision} + \text{Recall}}$.} providing a balanced metric that is particularly useful when the class distribution is imbalanced. Accuracy is the ratio of correctly predicted observations to the total observations,\footnote{$\text{Accuracy} = \frac{TP + TN}{TP + TN + FP + FN}$, where $TN$ is true negatives.} giving the overall effectiveness of the model in predicting both positive and negative cases.

Our findings show that using a DCI instance window size of 20 results in accuracy below 90\%. However, with a window size of over 40 DCI instances, accuracy surpasses 91\%. With over 100 instances, accuracy exceeds 96\%, and with over 140 instances, accuracy rises above 98\% for unseen datasets. As the number of DCI instances increases, accuracy continues to improve, as summarized in Table~\ref{tab:accuracy}, where precision, recall, and F1-score values represent the weighted average for each application. Despite this trend, continually increasing the DCI instance window size to improve accuracy is not practical, as it requires more time to collect the additional instances. Therefore, it is important to determine an optimal window size that balances accuracy with the time required for data collection. Section~\ref{subsec:findingwindowsize} explains the method for setting the appropriate DCI instance window size.

\begin{table}[t]
\label{tab:accuracy}
\centering
\caption{Performance Evaluation for Different DCI Instance Window Sizes}
\begin{tabular}{ccccccc}
\toprule
\textbf{Window Size} & \textbf{Precision} & \textbf{Recall} & \textbf{F1-score} & \textbf{Accuracy} \\
\midrule
20  & 0.878 & 0.877 & 0.876 & 0.877 \\
40  & 0.916 & 0.915 & 0.913 & 0.915 \\
60  & 0.940 & 0.939 & 0.939 & 0.939 \\
80  & 0.954 & 0.953 & 0.953 & 0.953 \\
100 & 0.967 & 0.965 & 0.965 & 0.965 \\
120 & 0.976 & 0.976 & 0.976 & 0.976 \\
140 & 0.981 & 0.980 & 0.980 & 0.980 \\
160 & 0.987 & 0.987 & 0.987 & 0.987 \\
\bottomrule
\end{tabular}
\vspace{0cm} 
\begin{flushright}
*Window Size = Number of DCI instances
\end{flushright}
\end{table}

\subsection{Evaluating Classification Time for Various DCI Instance Window Sizes}
\label{subsec:timerequired}
The frequency of DCI instance occurrences varies across different applications. VoIP call applications, which require real-time data transmission, and video streaming applications generally exhibit a relatively high frequency of DCI instances, although this can vary depending on the specific application. In contrast, music streaming applications tend to have a lower frequency of DCI instances. Consequently, even with the same DCI instance window size, the time required to gather these instances varies across applications.

Reducing the DCI instance window size allows for quicker classification but results in lower accuracy. Conversely, increasing the DCI instance window size improves accuracy but requires more time to gather the required DCI instances. 
Once the specified number of DCI instances is collected, classification can be performed almost immediately. Thus, the time taken to collect the DCI instances corresponding to the DCI instance window size effectively represents the time required for application classification. 

\begin{figure}[t]
\centering
\includegraphics[width=\columnwidth]{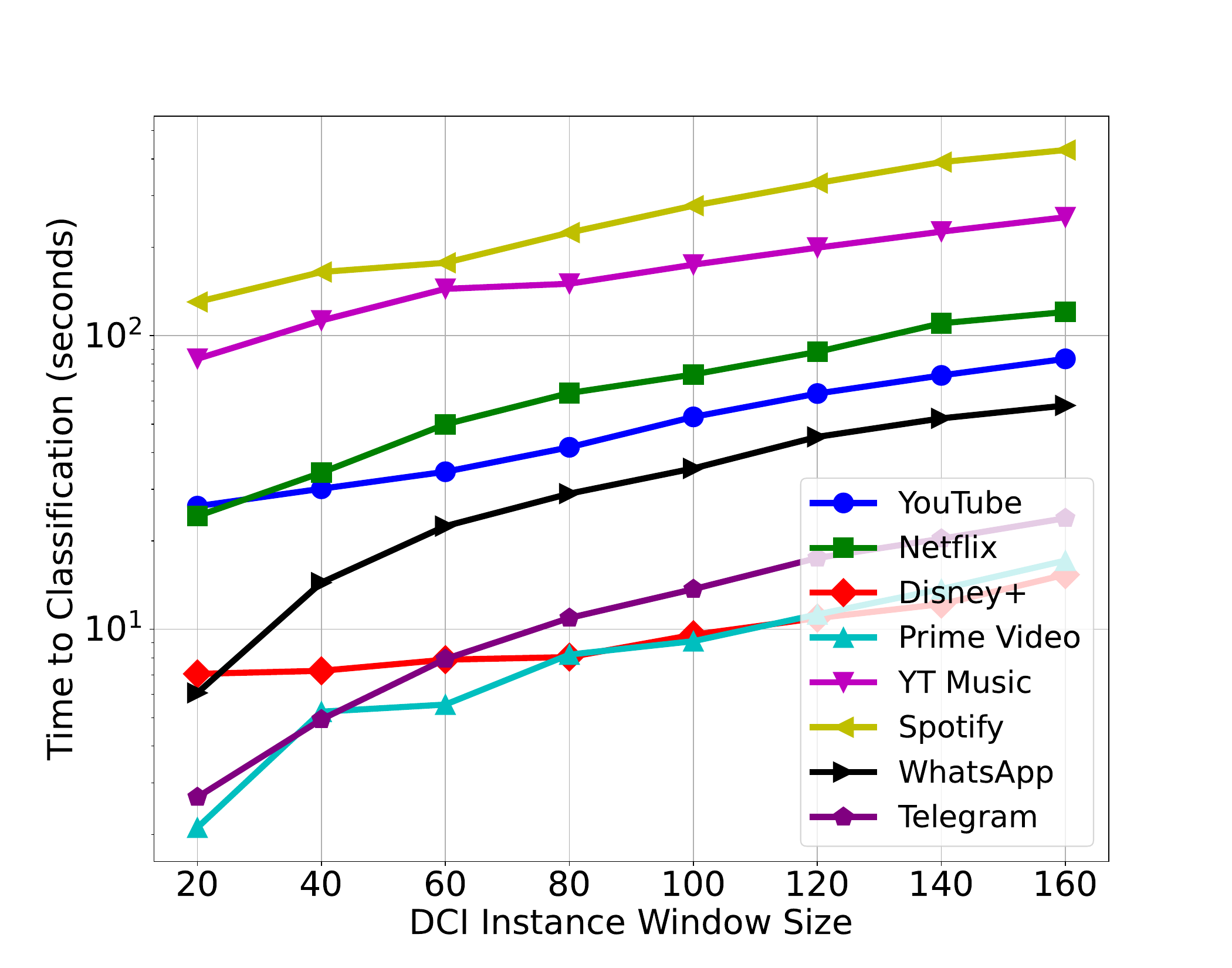}
\caption{Time to Application Classification in srsRAN 5G Testbed with the 5G sniffer}
\label{fig:timerequired}
\end{figure}

We measured the time required to classify each application using different DCI instance window sizes. The results are based on the average classification time across 50 to 100 evaluations per application. For a DCI instance window size of 40, which achieves an overall classification accuracy of approximately 91.5\%, the average classification times are as follows: YouTube takes about 26 seconds, Netflix about 24 seconds, Disney+ around 7 seconds, Prime Video approximately 2.1 seconds, YT Music around 83 seconds, Spotify about 130 seconds, WhatsApp around 6 seconds, and Telegram approximately 2.7 seconds. This demonstrates that application classification is performed adaptively based on the DCI instance generation rate of each application. The time required for classification with different DCI instance window sizes in our test environment for each application is shown in Figure~\ref{fig:timerequired}.

\subsection{Finding an Appropriate DCI Instance Window Size}
\label{subsec:findingwindowsize}
\begin{figure}[t]
\centering
\includegraphics[width=0.85\columnwidth]{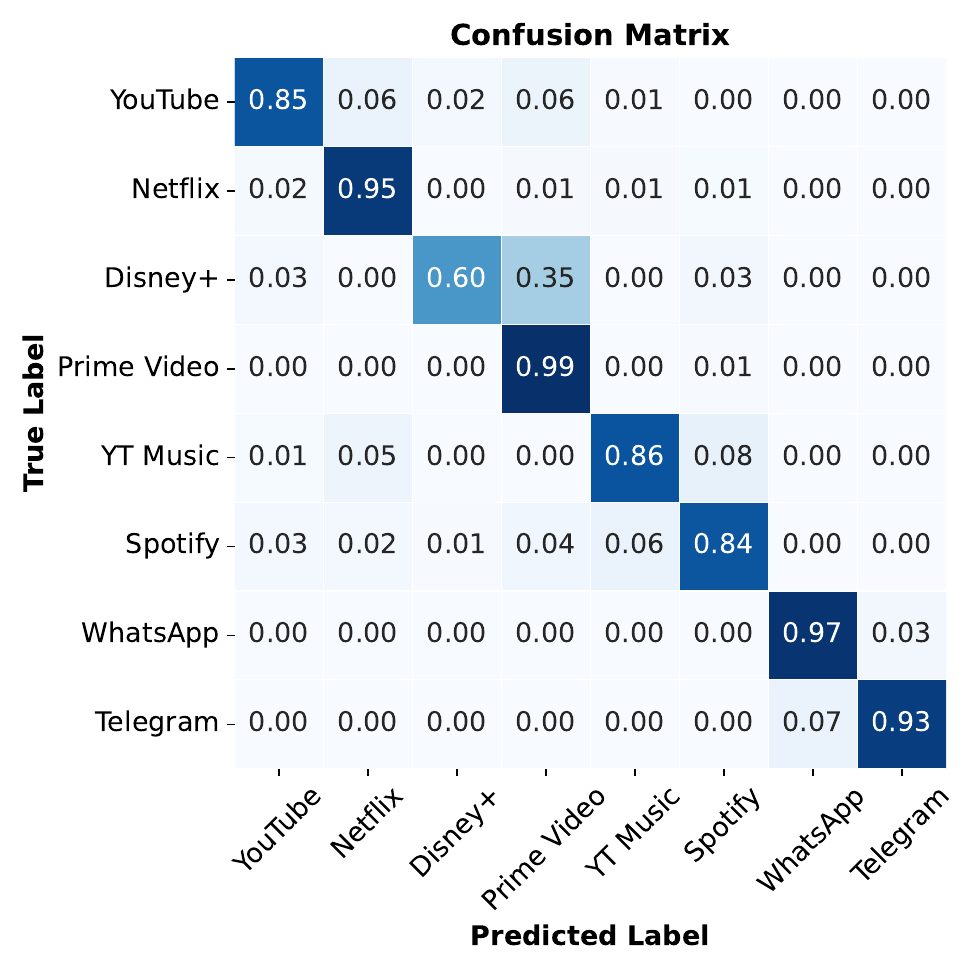}
\caption{Confusion Matrix with DCI Instance Window Size of 40 in srsRAN 5G Testbed}
\label{fig:confusion_40}
\end{figure}

The overall performance evaluation results are detailed in Section~\ref{subsec:classfication_accuracy}, with the classification accuracy for each application summarized as follows. When using a DCI instance window size of 40, the precision, recall, and F1 score are presented in Table~\ref{tab:accuracy_40}, and the corresponding confusion matrix is shown in Figure~\ref{fig:confusion_40}.

The confusion matrix is a tool used to visualize the performance of a classification algorithm. Each row represents the true class, while each column represents the predicted class. When expressed as percentages, the values in the confusion matrix indicate the proportion of correct predictions for each true class. For example, a value of 0.95 in the cell corresponding to the true class \textit{Netflix} and the predicted class \textit{Netflix} means that 95\% of the instances truly belonging to \textit{Netflix} were correctly predicted.

When analyzing the results with a DCI instance window size of 40, we find that VoIP call applications are classified with high accuracy, exceeding 93\%. However, the classification accuracy for video streaming and audio streaming applications is relatively lower. Therefore, if the goal is to classify VoIP call applications with more than 90\% accuracy, a DCI instance window size of 40 is sufficient. 
However, while it is possible to classify video streaming and audio streaming applications with a DCI instance window size of 40, a larger window size is required if the goal is to achieve classification accuracy above 90\%.

\begin{table}[t]

\centering
\caption{Performance Evaluation for each Application with DCI instance window size of 40 in srsRAN 5G Testbed}
\label{tab:accuracy_40}
\begin{tabular}{lccc}
\toprule
\textbf{Application} & \textbf{Precision} & \textbf{Recall} & \textbf{F1-score} \\
\midrule
\textbf{YouTube} & 0.933 & 0.847 & 0.888 \\
\textbf{Netflix} & 0.811 & 0.948 & 0.874 \\
\textbf{Disney+} & 0.868 & 0.595 & 0.706 \\
\textbf{Prime Video} & 0.910 & 0.988 & 0.947 \\
\textbf{YT Music} & 0.865 & 0.856 & 0.861 \\
\textbf{Spotify} & 0.920 & 0.843 & 0.880 \\
\textbf{WhatsApp} & 0.937 & 0.971 & 0.954 \\
\textbf{Telegram} & 0.966 & 0.934 & 0.950 \\
\midrule
\textbf{Accuracy} & & & 0.915 \\
\textbf{Macro avg} & 0.901 & 0.873 & 0.882 \\
\textbf{Weighted avg} & 0.916 & 0.915 & 0.913 \\
\bottomrule
\end{tabular}
\end{table}

\begin{figure}[t]
\centering
\includegraphics[width=0.85\columnwidth]{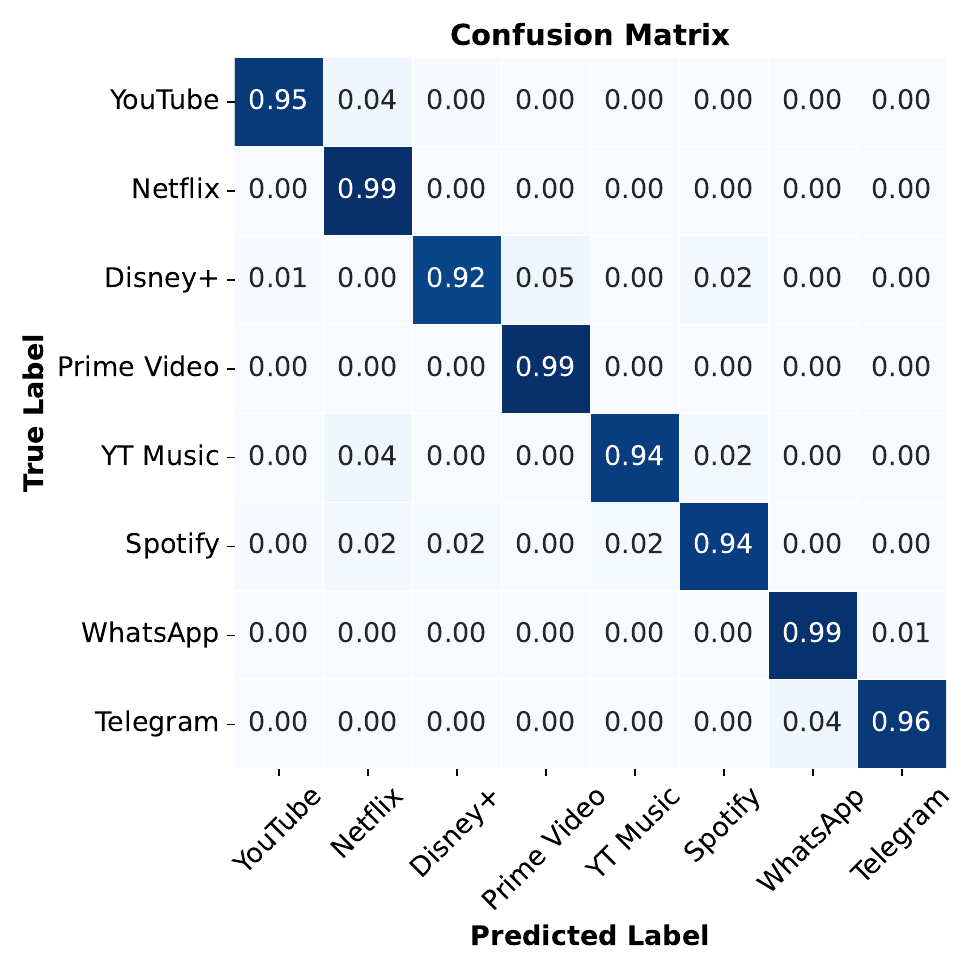}
\caption{Confusion Matrix with DCI Instance Window Size of 100 in srsRAN 5G Testbed}
\label{fig:confusion_100}
\end{figure}

In such cases, for example, when the DCI instance window size is set to 100, the results are shown in Table~\ref{tab:accuracy_100}, and the corresponding confusion matrix is illustrated in Figure~\ref{fig:confusion_100}. It is evident that increasing the window size significantly enhances the classification accuracy for video and audio streaming applications. Therefore, in these situations, setting the DCI instance window size to 100 or a level that meets the required accuracy would be appropriate. By balancing accuracy and classification time for different DCI window sizes, it is possible to determine an optimal DCI window size that fits the specific requirements of the situation.

\begin{table}[t]

\centering
\caption{Performance Evaluation for each Application with DCI instance window size of 100 in srsRAN 5G Testbed}
\label{tab:accuracy_100}
\begin{tabular}{lccc}
\toprule
\textbf{Application} & \textbf{Precision} & \textbf{Recall} & \textbf{F1-score} \\
\midrule
\textbf{YouTube} & 0.988 & 0.952 & 0.970 \\
\textbf{Netflix} & 0.849 & 0.994 & 0.916 \\
\textbf{Disney+} & 0.948 & 0.919 & 0.933 \\
\textbf{Prime Video} & 0.984 & 0.994 & 0.989 \\
\textbf{YT Music} & 0.974 & 0.940 & 0.957 \\
\textbf{Spotify} & 0.961 & 0.937 & 0.949 \\
\textbf{WhatsApp} & 0.963 & 0.986 & 0.974 \\
\textbf{Telegram} & 0.985 & 0.962 & 0.973 \\
\midrule
\textbf{Accuracy} & & &0.965 \\
\textbf{Macro avg} & 0.956 & 0.961 & 0.958 \\
\textbf{Weighted avg} & 0.967 & 0.965 & 0.965 \\
\bottomrule
\end{tabular}
\end{table}

\subsection{Evaluation in Alternative Commercial 5G Testbeds}
\label{subsec:evaluation_in_alternative_commercial_5g_testbeds}
To demonstrate that our algorithm is not limited to the 5G testbed composed of srsRAN and open5GS where we primarily conducted our experiments, but is also applicable in other environments, we conducted additional experiments using another commercial 5G testbed. The new commercial 5G testbed was configured using the core network and gNB provided by AMARI Callbox~\cite{amari_callbox}. In the 5G testbed using AMARI Callbox, approximately 10\% of the DCI instances generated by the gNB were successfully captured over the air by the 5G sniffer, resulting in a 10:1 ratio of captured DCI instances. Application classification was then performed using this limited data. This differs from the srsRAN and Open5GS setup, where only about 5\% of the DCI instances were captured and used for classification. The purpose of this comparison was to demonstrate that our algorithm could effectively classify applications in a different network environment, where up to 10\% of the DCI instances were captured, without requiring any modifications to the algorithm.

In the AMARI Callbox 5G testbed, we did not collect data for all applications previously tested, as the purpose was to demonstrate proof of concept. Instead, we collected data for representative applications for each service. For video streaming services, we collected data from YouTube and Netflix; for audio streaming services, we used YT Music; and for VoIP call services, we collected data from WhatsApp. We verified the performance of our proposed algorithm in distinguishing between these applications.

\begin{figure}[t]
\centering
\includegraphics[scale=0.53]{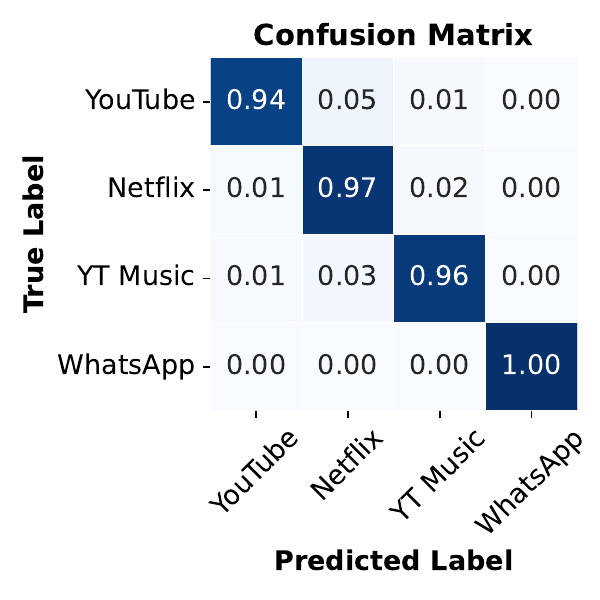}
\caption{Confusion Matrix with DCI Instance Window Size of 100 in AMARI 5G Testbed}
\label{fig:confusion_100_amari}
\end{figure}

To do this, we collected data for each application using the same method as described in Section~\ref{subsec:datacollection}, resulting in approximately 609 hours of data in total. This data was split into a 9:1 ratio, with 90\% used for training our deep learning model and 10\% used for validation. The DCI instance window size was set to 100. The results of these experiments are shown in Table~\ref{tab:amari}, and the corresponding confusion matrix is presented in Figure~\ref{fig:confusion_100_amari}. As the experimental results demonstrate, even with varying DCI instance capture rates across different networks, the algorithm can still achieve more than 97\% of accuracy in application classification if trained on the specific network, without requiring any modifications to the algorithm.

\begin{table}[t]
\centering
\caption{Performance Evaluation for DCI instance window size of 100 in AMARI 5G Testbed}
\label{tab:amari}
\begin{tabular}{lccc}
\toprule
 & \textbf{Precision} & \textbf{Recall} & \textbf{F1-score} \\
\midrule
\textbf{YouTube} & 0.982 & 0.936 & 0.958 \\
\textbf{Netflix} & 0.952 & 0.973 & 0.962 \\
\textbf{YT Music} & 0.936 & 0.959 & 0.947 \\
\textbf{WhatsApp} & 0.997 & 0.973 & 0.973 \\
\midrule
\textbf{Accuracy} & & &0.973 \\
\textbf{Macro avg} & 0.967 & 0.967 & 0.967 \\
\textbf{Weighted avg} & 0.973 & 0.973 & 0.973 \\
\bottomrule
\end{tabular}
\end{table}

\subsection{RNTI Acquisition}

Ludant \textit{et al.}~\cite{5gsniffing2023} identified the location of target users through the transmission of 5G traffic signatures, which enables the acquisition of the target user’s RNTI. We conducted experiments to obtain RNTIs using a method similar to the existing approach~\cite{5gsniffing2023}, utilizing a commercial testbed configuration. The RAN and core network were configured using the AMARI Callbox to set up a 5G SA network. Subsequently, 64 UEs were generated using the AMARI UE Simbox~\cite{amari_simbox} and connected to the AMARI Callbox base station via the radio interface. 
The target UE was also connected to the base station, and its RNTI was recorded using a diagnostic monitor tool~\cite{xcal}. The target UE was a Samsung Galaxy S24, which was connected to the base station via the radio interface.

\begin{figure}[t]
\centering
\includegraphics[width=\columnwidth]{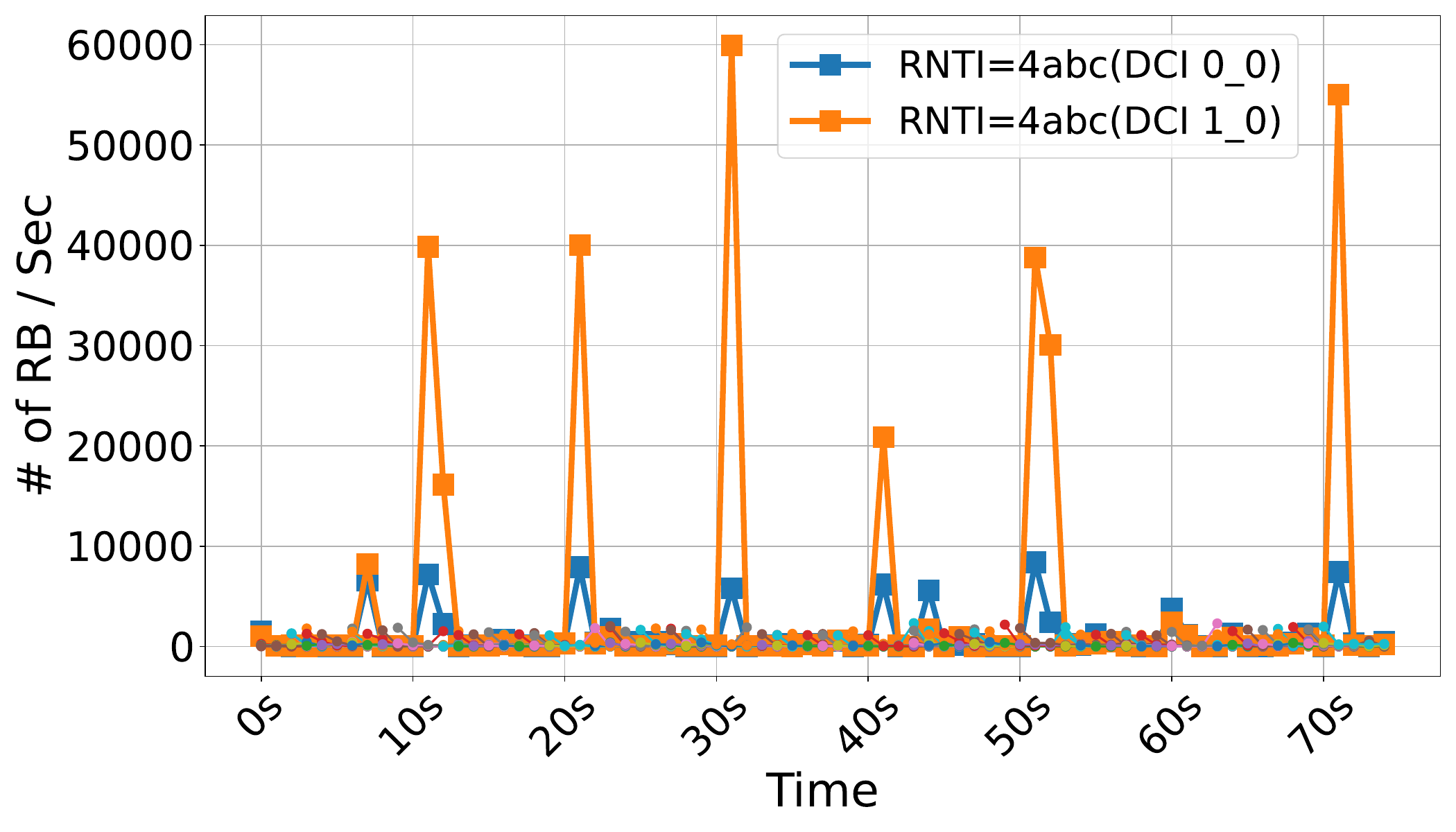}
\caption{Target RNTI Acquisition through Traffic Signature Generation}
\label{fig:rnti_acquisition}
\end{figure}

Traffic signatures can be crafted by an attacker in any desired form. Existing methods~\cite{5gsniffing2023} allow for data transmission through Signal and Telegram without triggering alerts for the target. However, to simplify the experiment, we designed a verification scenario where data is directly sent to the target to acquire the 5G RNTI. The procedure for acquiring the target RNTI is as follows: after transmitting more than 100 KB of data to the target UE, the RNTIs that are allocated resource blocks~(RB) above a certain threshold within the next 2 seconds of DCIs are extracted. The data transmission was repeated at least three times by sending a photo via Telegram. In our testbed, the threshold was set to 5,000 RBs per second for DCI 0\_0 and 10,000 RBs per second for DCI 1\_0. This threshold can be adjusted based on the amount of data sent or the test environment. Repeated data transmissions of this nature generate a form of traffic signature on the radio interface. Figure~\ref{fig:rnti_acquisition} shows the traffic pattern we created. We sent data five times at 10-second intervals and once at a 20-second interval. This method produces a unique pattern of five RB traffic bursts exceeding the threshold at 10-second intervals and one at 20-second intervals. Thus, RNTI \texttt{0x4ABC} indicates the target RNTI we are searching for. Even in an environment with heavy data usage by other users, an attacker can acquire the target’s RNTI by sending traffic signatures in this manner. As long as the RNTI receiving data at specific intervals set by the attacker remains unique, the RNTI can be acquired.

%% file: discussion.tex
In this section, we discuss the feasibility of 5G sniffing in real-world environments, assess the limitations of our study, and explore potential countermeasures against privacy attacks.

\subsection{Real-world 5G Sniffing}

To receive radio interface channel messages in cellular networks, specialized radio equipment such as Software-Defined Radios (SDRs) is necessary. Most studies, including ours, use the USRP B210 as the primary tool, although higher performance needs may require devices like the USRP X310~\cite{usrpx310}. Since 5G can support bandwidths exceeding 100 MHz, capturing signals from regions operating with such wide bandwidths necessitates the use of higher-performance SDRs like the USRP X310.

It is feasible to develop a 5G sniffer; however, open-source tools have been released later compared to LTE, and the control channel decoding process is somewhat more complex. As of July 2024, the only publicly available tool is the 5G Sniffer~\cite{5gsniffer}. This 5G Sniffer allows for 5G PDCCH decoding. To ensure stable operation of this tool, specific settings tailored to each testbed or operator are required, and these settings can be obtained through signaling related to the device's RRC setup. 

The description of this sniffer indicates that it currently supports only FDD mode. Therefore, in regions or countries where only Time Division Duplexing~(TDD) mode is used, 5G sniffing may be limited for now. However, once TDD mode is officially supported, the tool can be applied in these areas. In the United States, where operators like Verizon, AT\&T, and T-Mobile operate FDD mode within a 10MHz bandwidth, experiments using the USRP B210 may be conducted, provided ethical considerations and legal boundaries are followed.

\subsection{Limitation}
We demonstrated that fingerprinting of eight different mobile applications, such as YouTube, Netflix, and WhatsApp, is possible through 5G radio control channel decoding. However, the variety of applications registered on Google Play and the iOS App Store is much broader. As of July 30, 2024, there are approximately 2.33 million applications on Google Play and about 2 million on the iOS App Store, with around a thousand new applications being added daily~\cite{42matters}. Despite the vast number of applications, statistics show that users typically use only nine mobile applications per day and around 30 per month~\cite{dataprot, buildfire}. It can be inferred that meaningful application fingerprinting is possible if the focus is on high-traffic applications, such as video streaming services, social media, and messaging apps~\cite{dataprot_streaming,dataprot_social}. Our study demonstrated that applications with high market shares in video and music streaming and messaging categories can be distinguished. For other applications, it is believed that they can be adequately distinguished if sufficient training data is collected.

\subsection{Countermeasures}
One study~\cite{baek2023targeted} has proposed frequently updating the RNTI as a countermeasure, but this method may be ineffective against RNTI fixation attacks~\cite{oh2024enabling}. RNTI fixation attacks prevent the RNTI from changing by maintaining the radio connection. If a logic is added that changes the RNTI even while maintaining the radio connection, this could be a solution. However, this logic can only partially prevent targeted attacks and cannot prevent attacks such as cell scanning. Existing proposals~\cite{dubin2017know,schuster2017beauty, zhang2019statistical} to prevent fingerprinting by adding bitrate speed, data chunks, noise, and padding have also raised questions about their effectiveness at the service level, and they may lead to additional data charges. Ultimately, encrypting the DCI of the PDCCH appears to be a definitive solution, but it requires design changes to the encryption of the control plane. 
In-depth discussions are needed to develop solutions that improve operational efficiency.

%% file: conclusion.tex
Our proposed method not only pioneers as the first traffic classification approach on the 5G radio interface, but it also demonstrates remarkable robustness and high accuracy in application classification, even under conditions of significant PDCCH data loss within a 5G mobile network sniffing environment. 
 Furthermore, this was validated through experiments using both open-source and commercial 5G SA testbeds. These findings highlight vulnerabilities such as cell scanning and targeted privacy attacks. This paper shows that 5G SA networks are also susceptible to mobile activity fingerprinting and evaluates these vulnerabilities from multiple perspectives. Such evaluations are expected to prompt discussions on the protection of control channels in future 5G and 6G networks.